\documentclass[%
 aip,
 amsmath,amssymb,
 reprint,%
]{revtex4-1}

\usepackage{graphicx}
\usepackage{dcolumn}
\usepackage{bm}

\usepackage[utf8]{inputenc}
\usepackage[T1]{fontenc}
\usepackage{mathptmx}
\usepackage{etoolbox}
\usepackage{subfigure}
\usepackage{upgreek}
\usepackage{xcolor}

\makeatletter
\def\@email#1#2{%
 \endgroup
 \patchcmd{\titleblock@produce}
  {\frontmatter@RRAPformat}
  {\frontmatter@RRAPformat{\produce@RRAP{*#1\href{mailto:#2}{#2}}}\frontmatter@RRAPformat}
  {}{}
}%
\makeatother
\begin{document}
\preprint{AIP/123-QED}

\title[Appl.Phys.Lett DOI:10.1063/5.0137392]{Implementation of SNS thermometers into molecular devices for cryogenic thermoelectric experiments }
\author{Serhii Volosheniuk}
  \affiliation{%
Kavli Institute of Nanoscience, Delft University of Technology, Lorentzweg 1, 2628 CJ Delft, The Netherlands
}%

\author{Damian Bouwmeester }%
\affiliation{%
Kavli Institute of Nanoscience, Delft University of Technology, Lorentzweg 1, 2628 CJ Delft, The Netherlands
}%

\author{Chunwei Hsu}
\affiliation{%
Kavli Institute of Nanoscience, Delft University of Technology, Lorentzweg 1, 2628 CJ Delft, The Netherlands
}%

\author{H.S.J. van der Zant}
\affiliation{%
Kavli Institute of Nanoscience, Delft University of Technology, Lorentzweg 1, 2628 CJ Delft, The Netherlands
}%

\author{Pascal Gehring}
\email{pascal.gehring@uclouvain.be\\}
\affiliation {\small \textit IMCN/NAPS, Université Catholique de Louvain (UC Louvain), 1348 Louvain-la-Neuve, Belgium}
\date{\today}

\begin{abstract}
Thermocurrent flowing through a single-molecule device contains valuable information about the quantum properties of the molecular structure and in particular, on its electronic and phononic excitation spectra, and entropy. Furthermore, accessing the thermoelectric heat-to-charge conversion efficiency experimentally can help to select suitable molecules for future energy conversion devices, which --  predicted by theoretical studies -- could reach unprecedented efficiencies. However, one of the major challenges in quantifying thermocurrents in nanoscale devices is to determine the exact temperature bias applied to the junction.  
In this work, we have incorporated a superconductor-normal metal-superconductor (SNS) Josephson junction thermometer into a single-molecule device. The critical current of the Josephson junction depends accurately on minute changes of the electronic temperature in a wide temperature range from 100~mK to 1.6~K. Thus, we present a device architecture which can enable thermoelectric experiments on single molecules down to millikelvin temperatures with high precision.   
\end{abstract}

\maketitle

 Thermoelectric effects, i.e., the conversion between heat and charge currents, have received renewed interest from the nanoelectronics community \cite{Heremans2013,Gemma2021} \textcolor{black}{and in particular in the field of single-molecule electronics \cite{RincnGarca2016,Wang2019}.} Recent studies~\cite{Josefsson2018,Kleeorin2019,Dutta2019,entropy,Kondopaper} show that thermoelectric measurements at cryogenic temperatures contain important information about the physical and quantum-thermodynamic properties of nanoscale \textcolor{black}{and molecular} systems, which are hard to access with conventional transport experiments. Furthermore, according to theoretical predictions, molecular devices could achieve a very high dimensionless figure of merit~\cite{molecularheatmonsters} $ZT$  -- a quantity that is a measure for the heat-to-charge conversion efficiency. While the highest observed $ZT$ of inorganic materials is currently only~\cite{BabaNature} about 5-6, the predicted $ZT$ of molecular heat engines could reach values of $ \approx 100$, which would result in efficiencies close to the Carnot efficiency limit~\cite{Lambert}. This efficiency could be even further enhanced at cryogenic temperature~\cite{Achim}. However, to quantify thermoelectric effects it is primordial to know the exact temperature drop across a molecular junction. This is in particular challenging at cryogenic temperature because most thermometers used so far are metal-based resistive sensors with low sensitivity at $T < 20$~K. This work  fills this gap by developing a superconducting thermometer, which is sensitive down to mK temperature, and by implementing this thermometer into an electromigrated break junction (EMBJ) device.
 
Different types of low-temperature thermometry approaches have been explored in the literature, including Johnson Noise thermometry~\cite{Noise}, Coulomb Blockade thermometry~\cite{PekolaCBTarray}, and Hybrid Tunnel Junction thermometry~\cite{oldNIS}. In this article, we focus on superconductor-normal metal-superconductor (SNS) thermometers because of their excellent properties: low impedance, high sensitivity at low temperatures, and a negligible access resistance~\cite{duttathesis}. Furthermore, they can be easily implemented into molecular junctions: To this end, we \textcolor{black}{further improve} our single-molecule thermoelectric junctions~\cite{Gehring2019Efficientheating}, by contacting the source and drain gold contacts with two superconductors, forming a local SNS junction. By measuring the critical current over this SNS junction the temperature can be extracted. In this manner, the contacts can be simultaneously used for thermometry and transport measurements. Here, we use a molybdenum rhenium (MoRe) superconductor, for its high critical temperature of 8.7~K (with a zero-temperature superconducting gap of about $\Delta\approx 1.8 k_{\rm B}T_{\rm C}\approx$ 1.3~meV).

\begin{figure}
\includegraphics[width=1.0\linewidth]{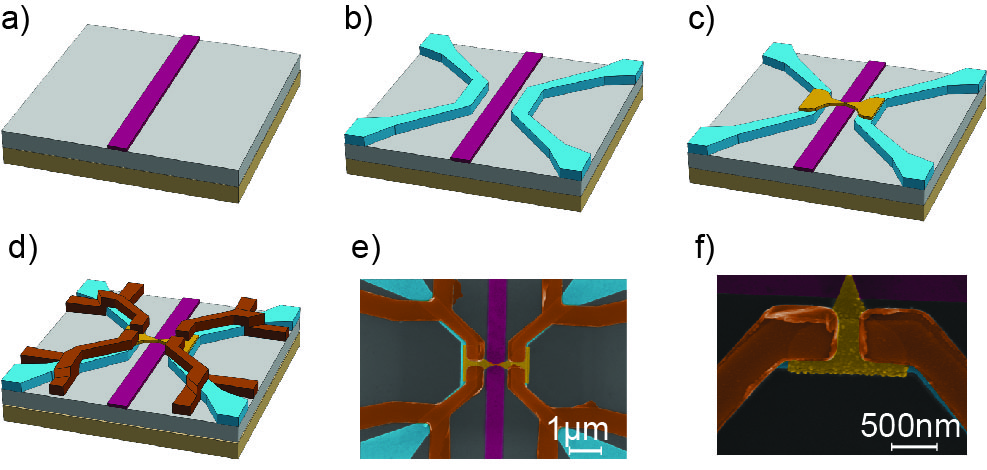}
\caption{\label{fig:epsart} (a)-(d) Fabrication process overview. (a) Electron beam (E-Beam) deposition of the local back gate electrode (purple). (b) Deposition of sample heaters (blue). (c) Deposition of  13~nm thick golden bridge (yellow). (d) Metal sputtering of superconducting contacts which act as source/drain connections and thermometers. (e)-(f) False-colored scanning-electron microscopy (SEM) images of the fabricated sample. (e) False-colored SEM picture of the thermopower device and (f) of the MoRe/Au/MoRe junction; purple: gate electrode; blue: heaters; yellow: gold bridge; brown: MoRe contacts.}
\label{fig:Fig1}
\end{figure}

 The fabrication procedure of the devices on Si wafers with a 285~nm layer of SiO$_2$ oxide is illustrated in Figures~\ref{fig:Fig1}(a)-(d).
 First (Figure~\ref{fig:Fig1}(a)), the local gate was made by depositing a 1~nm thick adhesion layer of titanium (Ti) and a 7~nm thick layer of palladium (Pd) by standard electron-beam lithography and metal evaporation. This gate thickness is chosen to decrease the heat transport from source to drain~\cite{Gehring2019Efficientheating}. Then, the heater was fabricated by depositing 3~nm of Ti and 27~nm of Pd (Figure~\ref{fig:Fig1}(b)). Subsequently, 10~nm Al$_2$O$_3$ was deposited by atomic layer deposition. The aluminium oxide  forms the insulating layer between the heater and the electrical contacts, and acts as the dielectric of the local gate. Afterwards, the 13~nm thick Au bridge was made (see Figure~\ref{fig:Fig1}(c)). To get a high-quality gold bridge, the deposition rate during evaporation was kept low ($0.5$~A/s) and a high vacuum around $10^{-8}$ mbar was used. In the last step, the 100~nm thick MoRe contacts were created by electron-beam lithography, metal sputtering and lift-off (Figure~\ref{fig:Fig1}(d)).
 
 False-colored scanning electron microscope (SEM) images of the final device together with a zoom-in of an individual Josephson junction are shown in Figures~\ref{fig:Fig1}(e)-(f). The spacing between the superconducting contacts varies, and is 253~nm on one side of the gold bridge (thermometer A) and 247~nm on the other side (thermometer B), in the device shown here.
 
\begin{figure}[!ht]
\includegraphics[width=1.0\linewidth]{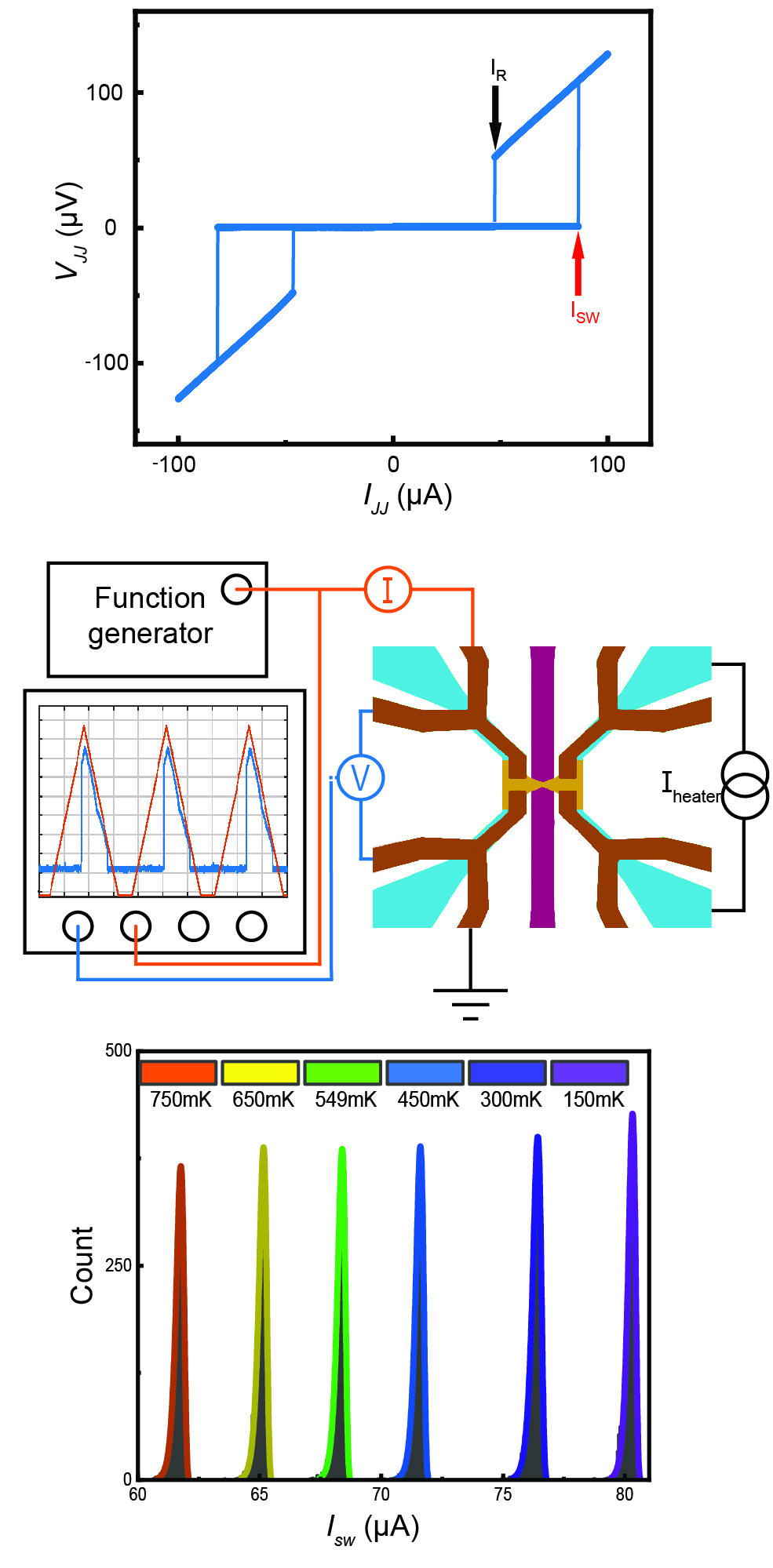}
\caption{\label{fig:epsart2} a) Voltage vs. current characteristics of a MoRe/Au/MoRe junction. The current was ramped up from 0 to 100~$\upmu$A, then to -100~$\upmu$A and back to zero. b) Schematic of the signal path during an AC experiment. A function generator was used to bias the superconducting thermometer with a current $I$ and to measure the voltage drop $V$ on the thermometer. A DC heater current $I_{\rm heat}$ can be applied to one of the micro heaters to generate a thermal bias on the junction.  c) Histogram of stochastic switching current $I_{\rm sw}$ of a SNS junction at different temperatures.}
\label{fig:fig2}
\end{figure}

 To calibrate the thermometers, devices were cooled down to 100~mK in a dilution refrigerator. A four-point measurement scheme was applied, where a DC current ($I_\mathrm{JJ}$) was biased over the junction and the voltage response ($V_\mathrm{JJ}$) was simultaneously recorded. A typical DC current-voltage characteristic at 100~mK for thermometer A is shown in Figure~\ref{fig:fig2}(a). In the low-current regime, the gold bridge between the two superconductors is proximitized and the Cooper pairs can move from one superconductor to the other without dissipation, forming an Andreev bound state~\cite{Abound}. At a certain current level -- i.e., at the switching current ($I_{\rm SW}$) -- the gold weak link changes its state from superconducting to normal, resulting in a voltage increase in the current-voltage ($I-V$) characteristic. From the slope of the $I-V$ characteristic at $I > I_{\rm SW}$ the normal resistance of the gold weak link is calculated and the diffusion coefficient of the electrons in this region is estimated. When ramping the current back, the gold weak link becomes superconducting again at the retrapping current ($I_{\rm R}$). This current value typically differs from $I_{\rm SW}$, which has been explained by capacitive effects in the junction~\cite{StewartW} or by heating of the electrons in the normal conducting junction during current sweeping~\cite{CourtoisHysteresis}. Since the geometrical capacitance of the junctions used in this work is not sufficiently large to explain the existence of the retrapping current, we attribute the observed hysteresis to heating.

Importantly, the switching to the superconducting state with increasing current bias is a stochastic process~\cite{AngersDC}, so several $I-V$ curves need to be recorded and a mean value of $I_\mathrm{SW}$ is calculated. To this end, the critical current was measured with an AC technique~\cite{WinkelmanDutta} (see Figure~\ref{fig:fig2}b). In a typical measurement the junction is biased with a 300~Hz triangular AC current from -20~$\upmu$A to 87~$\upmu$A. At these settings, gold stays in the normal state only for a short time, preventing the system from heating. Also, the offset is chosen to ensure that the junction stays proximitized in the negative current part and can relax to the base temperature. Using this approach, in a typical experiment we recorded 3000 measurements of switching events with a current resolution of about 0.07~$\upmu A$. 

\begin{figure}[!ht]
\includegraphics[width=1.0\linewidth]{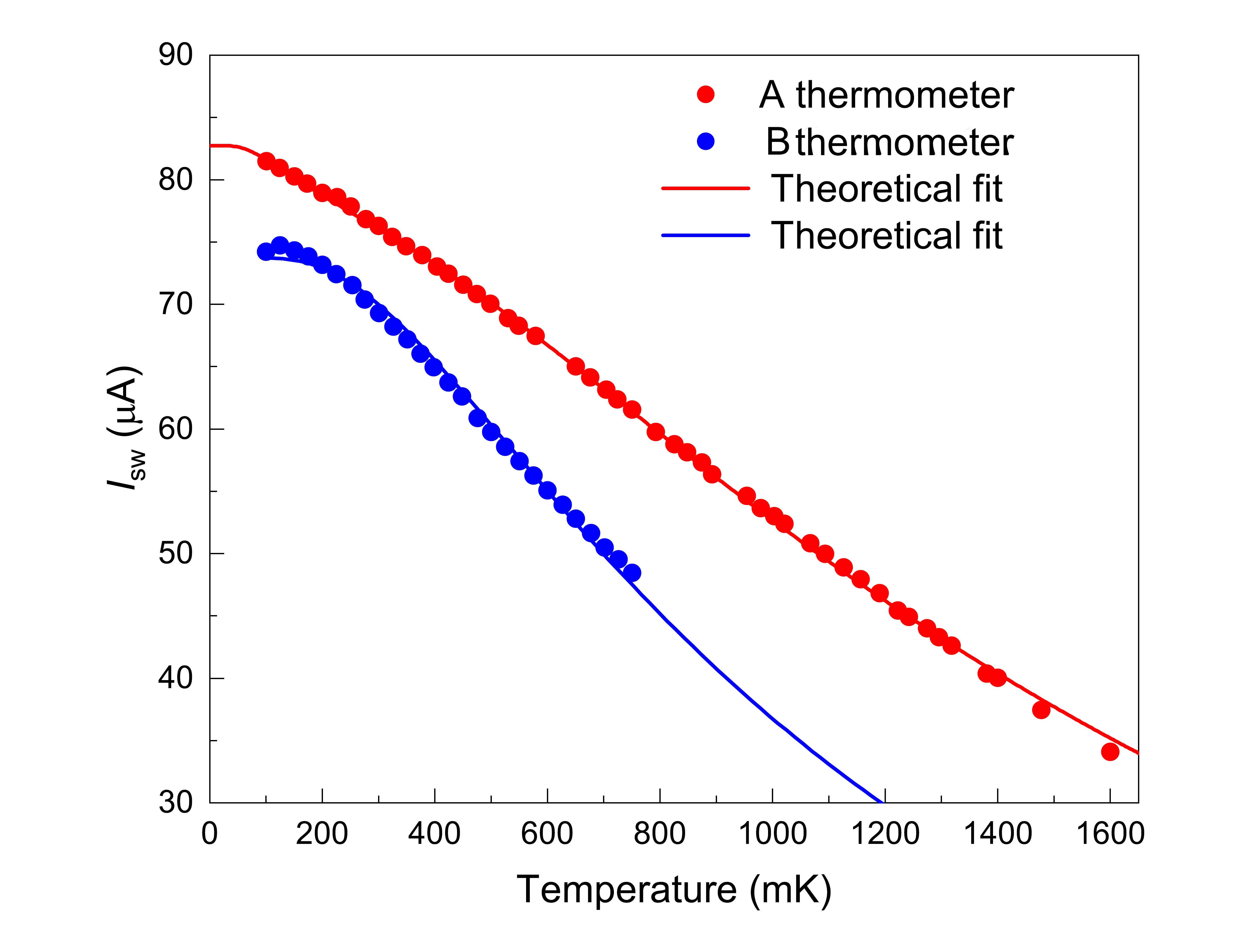}
\caption{\label{fig:fig3} Temperature dependence of the averaged switching current, $<I_{\rm SW}>$, for thermometers A (red) and B (blue), respectively. Theoretical fits using the Usadel equation (see main text) are shown as solid lines.
}
\end{figure}

Using the AC measurement technique, it is possible to measure $I_{\rm SW}$ at different base temperatures and calibrate the thermometers in the device. The $I_{\rm SW}$ histograms for thermometer A are shown in Figure~\ref{fig:fig2}(c). It can be seen that the switching current is very sensitive to the sample temperature and decreases from 80~$\upmu$A at 150~mK to 61~$\upmu$A at 750~mK. We also note an asymmetry in the switching current distributions. Such asymmetry has been observed  before~\cite{Spahr} and attributed to thermal fluctuations. \textcolor{black}{The stochasticity of switching current is dominated by quantum noise at temperatures below the Thouless energy, while thermal fluctuations become the dominating mechanism at higher temperatures (see discussion in SI III).} From the fits to the current histograms we extract the average switching current $<I_\mathrm{sw}>$ as a function of temperature for the two thermometers (see Figure~\ref{fig:fig3}). The sensitivity of the thermometers, defined as $\frac{dI_{\rm SW}}{dT}$, in the temperature range from 100~mK to 750~mK is 31.5~$\mu$A/K and 42~$\mu$A/K for thermometer A and B, respectively \textcolor{black}{(see SI II. for data of other thermometers and SI V. for a comparison with other Au-based SNS thermometers).} These values are typical for long SNS junctions and similar to those reported for Nb-Cu-Nb SNS junctions~\cite{dubos1}. \textcolor{black}{Furthermore, both thermometers have a temperature resolution better than 15~mK within the investigated temperature range (see SI IV. for more details).} 

In the following, we discuss the temperature dependence of $<I_\mathrm{sw}>$ (see Figure~\ref{fig:fig3}). In Josephson junctions this dependence is very sensitive to the interplay between the  energy scales of the proximity effect, the Thouless energy, $E_{\rm TH}=\frac{\hbar D}{L^2}$, and the superconducting gap $\Delta$. Here, $D=\frac{v_{\rm F} l_{\rm e}}{3}$ is the diffusion coefficient in the normal metal, where $v_{\rm F},~l_{\rm e}$ are the Fermi velocity and the elastic mean free path of electrons, respectively, and $L$ is the distance between the superconducting electrodes. In the long junction limit $\Delta>E_{\rm TH}$, the temperature dependence of $I_{\rm SW}$ can be described theoretically by using a quasi-classical approach based on Green's functions in imaginary space~\cite{Zaikin1,Zaikin2}; it requires solving the Usadel equation at all energies~\cite{Zaikin2}.

We used the~\textit{usadel1} package~\cite{virtanen} to fit the experimental data. We assumed that the interfaces between the superconductor and the normal metal are perfectly transparent and that the phase difference between the superconductors is fixed at $\phi=1.27\pi/2$, similar to Dubos \textit{et al.}~\cite{Zaikin2}. Furthermore, we fixed the normal resistance, $R_{\rm N}$, of the junction and $\Delta$ of the superconductor to values obtained from transport measurements. The solid lines in Figure~\ref{fig:fig3} show the resulting fits where the Thouless energy and the suppression coefficient, $\alpha$, which accounts for the non-ideality of the normal metal-superconductor contact, were used as fitting parameters. Details about the fitting and the resulting parameters can be found in the supporting information.  
We find a suppression coefficient $\alpha \approx 0.5$, similar to the value obtained by Courtois~\textit{et.al.}~\cite{CourtoisHysteresis}. The Thouless energies, $E^{\rm FIT}_{\rm TH} = 49.4 - 87.2 \mu$eV, obtained from the fits in Figure~\ref{fig:fig3} are close to the values determined from the transport measurements, $E^{\rm EXP}_{\rm TH} \approx 40 \upmu$eV, which can be calculated as~\cite{Janssen2001} $E^{\rm EXP}_{\rm TH}=\frac{1}{V N(E_{\rm F})}\frac{R_{\rm Q}}{2 \pi R_{\rm N} }$, where $V=L\cdot w \cdot t$ is the volume of the gold film, $N(E_{\rm F})$ is the volumetric electronic density of states of gold at the Fermi level and $R_{\rm Q}=\frac{h}{2e^2}$ is the resistance quantum. $N(E_{\rm F})$ can be obtained from the electronic specific heat\cite{Stewart1983} $\gamma = 0.69 \cdot 10^{-3}$ $\rm J\ mol^{-1} \ K^{-2}$ and the molar volume\cite{Singman1984} $v = 1.021 \cdot 10^{-5}$ $\rm m^3\ mol^{-1}$ as\cite{Beck1970} $N(E_{\rm F})=\frac{3\gamma}{\pi^2 k_{\rm B}^2 v}=1.73 \cdot 10^{28} $ $\rm eV^{-1}\  m^{-3}$.

In summary, we implemented an SNS superconducting thermometer in a molecular thermoelectric device. MoRe is used as the superconductor, which allows to perform thermometry in a temperature range from 100~mK to 1.6~K with a high sensitivity of up to $40~\upmu$A/K. Other than previously used resistance thermometers, the SNS thermometers developed in this work directly measure the temperature of the electronic system in the immediate proximity to the molecule. Therefore, our devices will allow to e.g. extract the absolute Seebeck coefficient from thermopower measurements on a molecule and will thus pave the way for precise investigations of molecular heat engines.




\section*{supplementary materials}
\textcolor{black}{See supplementary materials SI I. for theoretical fit to thermometer A with two parameters; SI II. for results of test thermometers with different L-- spacing; SI III. for discussion of $I_{\rm SW} $ fluctuations; SI IV. for temperature resolution of thermometers; and SI V. for comparison with other Au-based SNS junctions.}

\begin{acknowledgments}
The authors acknowledge financial support from the F.R.S.-FNRS of Belgium (FNRS-CQ-1.C044.21-SMARD, FNRS-CDR-J.0068.21-SMARD, FNRS-MIS-F.4523.22-TopoBrain), from the Federation Wallonie-Bruxelles through the ARC Grant No. 21/26-116 and from the EU (ERC-StG-10104144-MOUNTAIN, FET-767187-QuIET). This project (40007563-CONNECT) has received funding from the FWO and F.R.S.-FNRS under the Excellence of Science (EOS) programme. This work was supported by the Netherlands Organisation for Scientific Research (NWO/OCW), as part of the Frontiers of Nanoscience program and Natuurkunde Vrije Programma's: 680.90.18.01.
\end{acknowledgments}
\section*{Author Declarations}
\subsection*{Conflict of Interest}

    The authors have no conflicts to disclose. 

\section*{Author Contributions}

{\bf{Serhii Volosheniuk:}} Investigation (lead);  Resources (lead); Methodology (lead); Visualization (lead) Software (equal); Validation (equal); Writing - original draft (lead); Writing review \& editing (equal).
{\bf{Damian Bouwmeester:}} Validation (equal); Software (equal); Investigation (support).
{\bf{Chunwei Hsu:}} Validation (equal); Methodology (support); Resources (support).
{\bf{H.S.J. van der Zant:}} Supervision (equal); Writing - review \& editing (equal)
{\bf{Pascal Gehring:}} Project administration (lead); Supervision (equal); Writing -review \& editing (equal)
\section*{Data Availability Statement}

The data that support the findings of this study are available from the corresponding author upon  request.

\section*{References}
\bibliography{aipsamp}

\end{document}


\preprint{AIP/123-QED}

\title[SI]{Supplementary material for Implementation of SNS thermometers into molecular devices for cryogenic thermoelectric experiments }
\author{Serhii Volosheniuk}
  \affiliation{%
Kavli Institute of Nanoscience, Delft University of Technology, Lorentzweg 1, 2628 CJ Delft, The Netherlands
}%

\author{Damian Bouwmeester }%
\affiliation{%
Kavli Institute of Nanoscience, Delft University of Technology, Lorentzweg 1, 2628 CJ Delft, The Netherlands
}%

\author{Chunwei Hsu}
\affiliation{%
Kavli Institute of Nanoscience, Delft University of Technology, Lorentzweg 1, 2628 CJ Delft, The Netherlands
}%

\author{H.S.J. van der Zant}
\affiliation{%
Kavli Institute of Nanoscience, Delft University of Technology, Lorentzweg 1, 2628 CJ Delft, The Netherlands
}%

\author{Pascal Gehring}
\email{Email: pascal.gehring@uclouvain.be\\}
\affiliation {\small \textit IMCN/NAPS, Université Catholique de Louvain (UCLouvain), 1348 Louvain-la-Neuve, Belgium}

\maketitle
\section{Thermometer A: fits and discussion}
\begin{table}[h]
\centering
\begin{tabular}{ |c|c|c|c|c|c|c| } 
 \hline
 Thermometer & $w$ (nm) & $L$ (nm) & $R_{\rm N}$ ($\Omega$)& $E^{\rm EXP}_{\rm TH}$ ($\rm \mu \rm eV$)  &$ E^{\rm FIT}_{\rm TH}$ ($\rm \mu \rm eV$)& $\alpha$  \\ 
 \hline
  A& 511 & 253 & 1.7 &43.0 & 49.4 &0.31\\ 
   A$^*$& 511 & 253 & 1.7 &43.0 & 87.2/7.5 &0.5\\ 
   B & 503 & 247& 1.8 & 40.9  &31.5 &0.47\\
 \hline
\end{tabular}
\caption{Sample parameters. The junctions width ($w$), length ($L$), normal resistance ($R_{\rm N}$) are listed.  The experimental Thouless energy ($E^{\rm EXP}_{\rm TH}$) is calculated from the normal resistance and geometry of the sample. The Thouless energy, $E^{\rm FIT}_{\rm TH}$, and suppression coefficient, $\alpha$ are determined from the fit to the data. (Note that the A$^*$ and B fits are shown in the main text.) }
\label{table:1}
 
\end{table}
\begin{figure}[!ht]
\includegraphics[width=1.0\linewidth]{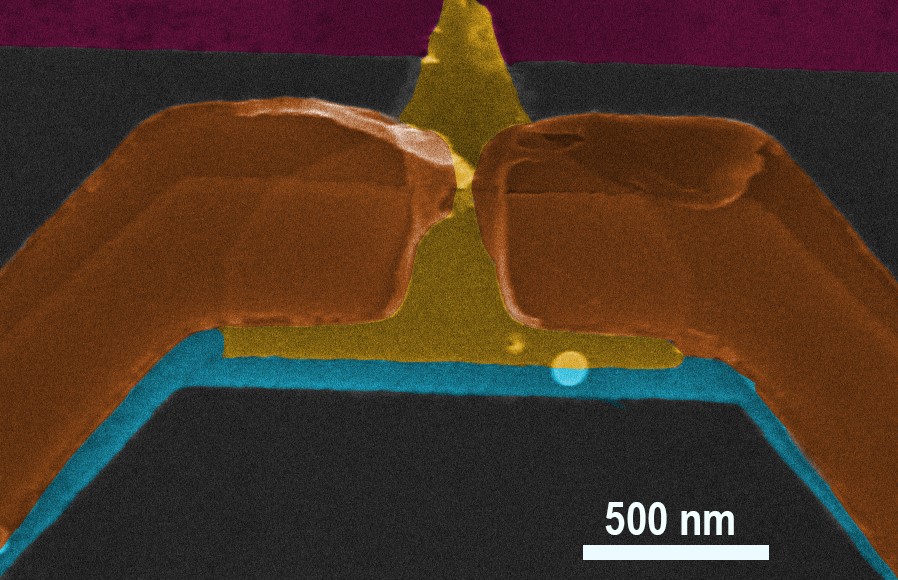}
\caption{\label{fig:fig1}  False-colored SEM picture of the A thermometer; purple: gate electrode; blue: heaters; yellow: gold bridge; brown: MoRe contacts.
}
\end{figure}
\begin{figure}[!ht]
\includegraphics[width=1.0\linewidth]{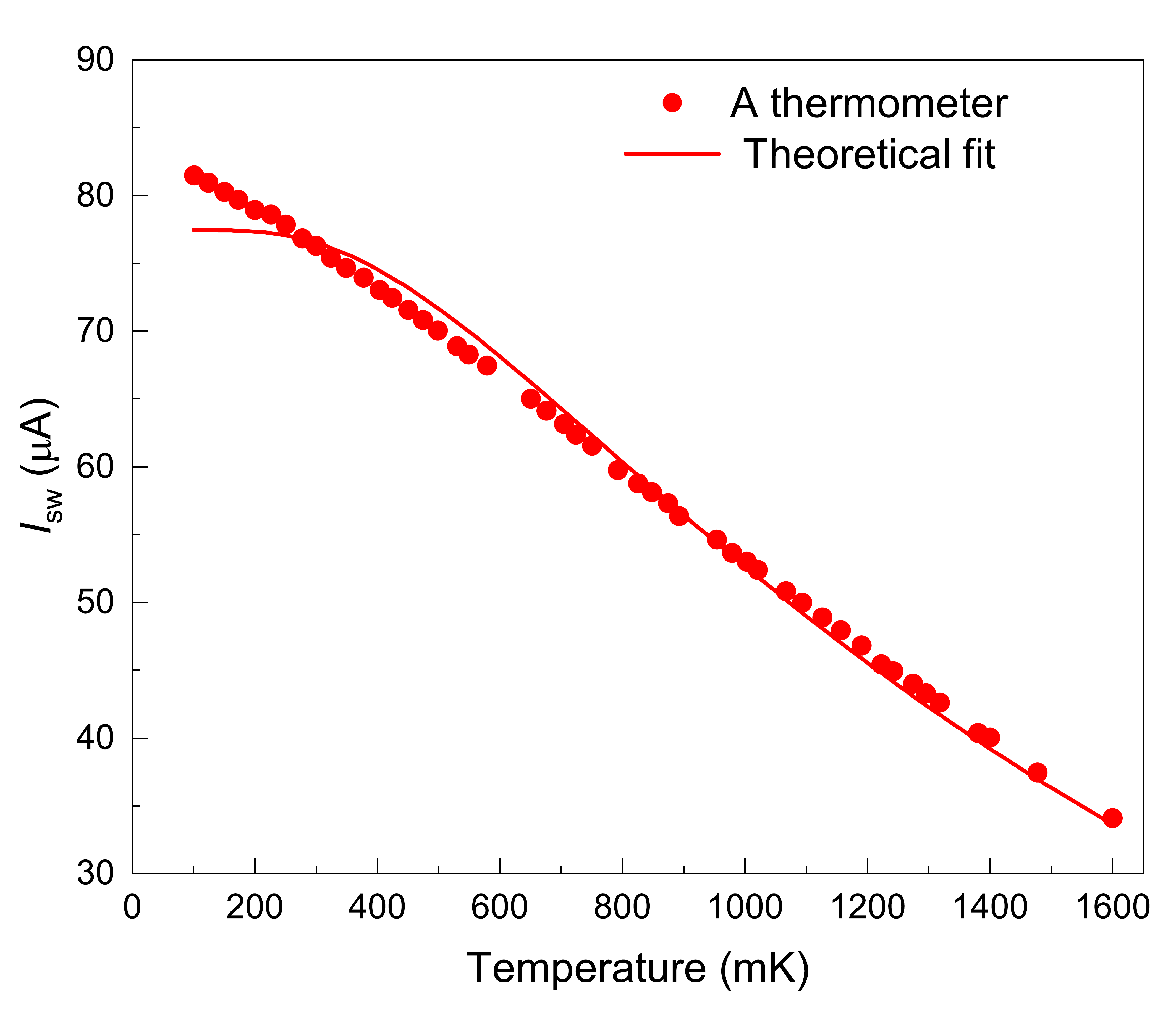}
\caption{\label{fig:fig3} Temperature dependence of the averaged switching current, $<I_{\rm SW}>$, for thermometer A and a theoretical fit to the data using the Usadel equation with two parameters $\alpha$ and $E_{\rm TH}^{\rm FIT}$ are shown as solid lines.
}
\end{figure}

The best theoretical fit with two parameters $\alpha$ and $E_{\rm TH}^{\rm EXP}$ is shown on Figure~\ref{fig:fig3}. A clear disagreement in the lower temperature range is present that could be related to the fabrication issues. It is possible that edge sputtering artefacts are touching the golden bridge and that the spacing between the MoRe contacts is not homogeneous (see Figure~\ref{fig:fig1}). In this case, the 1D approximation that we use would not be valid and 2D calculations would be required. To avoid complex calculations we decided to focus on a qualitative comparison and simulated the junction with two junctions connected in parallel with $ I_{\rm SW}= I_{\rm SW1}+ I_{\rm SW2}$, and $ R=\frac{ R_1R_2}{ R_1+R_2}$. This approach allows us to get a good fit for the data (see main text for the fit, and fitting parameters in Table~1). The extracted high Thouless energy for the first junction is two times smaller than the experimental value and corresponds to the part of the thermometer where edge sputtering artefacts are touching gold, thus making the effective distance smaller. The existence of a second Thouless energy $E_{\rm TH}=7.5 \mu$eV can possibly be explained by a discontinuity of the gold film at the step edge over the heater electrode. The heater electrode is $30$ nm thick, which is considerably thicker than the $13$ nm thick gold film. Thus a parallel SNS junction with a relatively larger volume and normal state resistance, thus smaller Thouless energy, may be formed by the EMBJ constriction.
\newpage
\section{Results of other thermometers}
To optimize thermometer design we varied the spacing between superconducting contacts. In Fig. S.3. we present the switching current and theoretical fits to the data for two test samples C and D with $L$ of 300~nm and 350~nm, respectively. The parameters of the junctions and estimated theoretical fit values are listed in the Table SII. We do observe at low temperatures that $I_{\rm SW}$ decreases with increasing length of the normal part of the junction. 
\begin{table}[h]
\centering
\begin{tabular}{ |c|c|c|c|c|c|c|c| } 
 \hline
 & $w$ (nm) & $L$ (nm) &$t$ (nm)& $R_{\rm N}$ ($\Omega$)& $E^{\rm EXP}_{\rm TH}$ ($\rm \mu \rm eV$)  &$ E^{\rm FIT}_{\rm TH}$ ($\rm \mu \rm eV$)& $\alpha$  \\ 
 \hline
   C& 440 & 305 &14& 1.6 &38.4 &23.1  &0.45\\
   D& 436 & 332 &14& 1.5 &36.5 & 25.8 &0.31\\
   
 \hline
\end{tabular}
\caption{Sample parameters for thermometers C and D. The junctions width ($w$), length ($L$), normal resistance ($R_{\rm N}$) are listed.  The experimental Thouless energy ($E^{\rm EXP}_{\rm TH}$) is calculated from the normal resistance and geometry of the sample. The Thouless energy, $E^{\rm FIT}_{\rm TH}$, and suppression coefficient, $\alpha$, are determined from the fit to the data. }
\label{table:2}
 
\end{table}

\begin{figure}[!ht]
\includegraphics[width=1.0\linewidth]{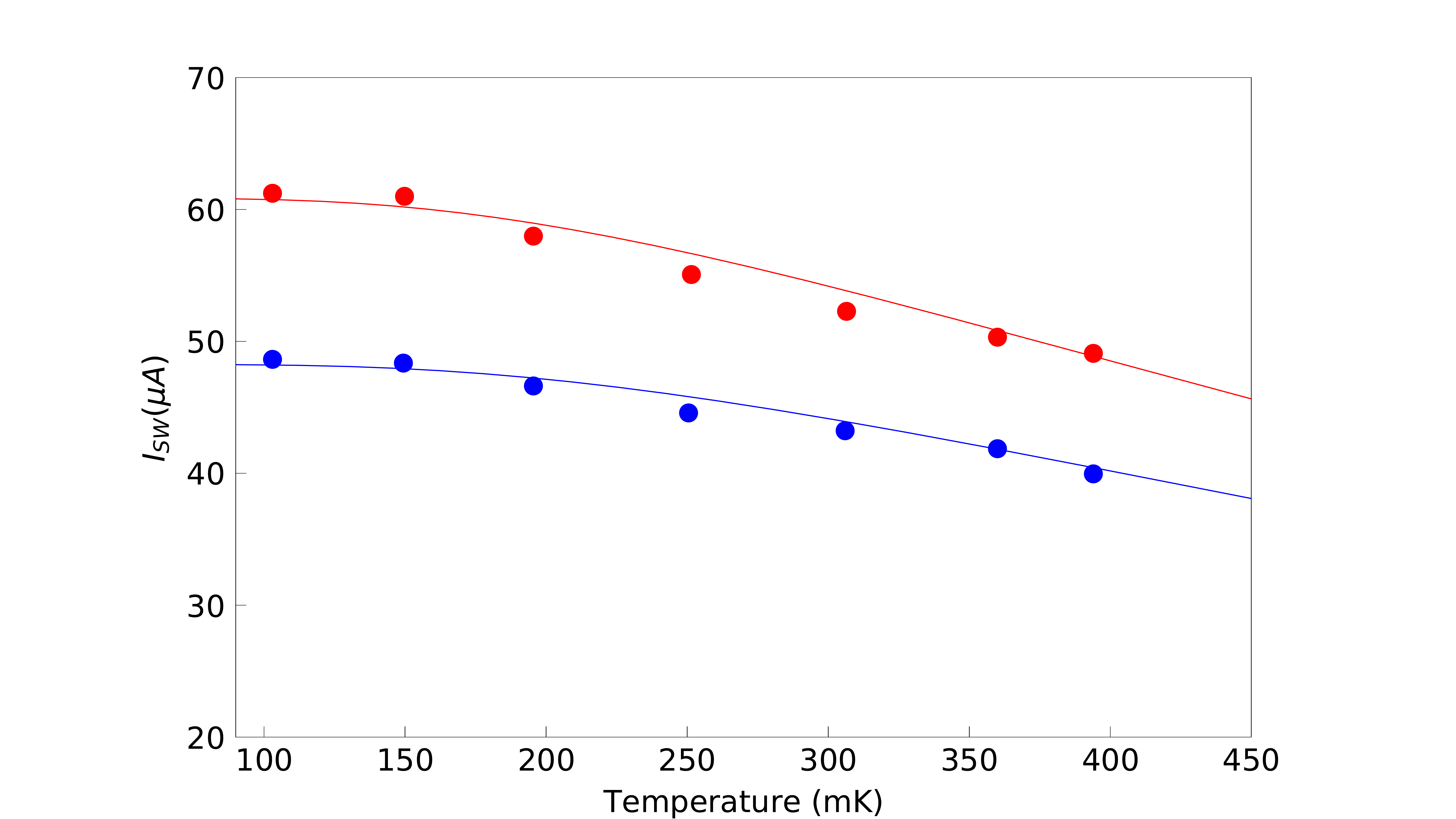}
\caption{\label{fig:fig3} Temperature dependence of the averaged switching current, $<I_{\rm SW}>$, for thermometers  C (red data points), and D (blue data points). Theoretical fits using the Usadel equation are shown as solid lines. 
}
\end{figure}

\newpage

\section{Origin of the fluctuations of switching current}

The JJ switching is a stochastic process, and the distribution of switching currents is influenced by thermal activation and quantum noise/tunneling. The standard deviation normalized by the switching current, $\frac{\sigma}{I_{\rm SW}}$, as a function of temperature is shown in Fig. S4. 
In Fig. S4, there are two different regions: 1) $\frac{\sigma}{I_{\rm SW}}$ stays constant at the low temperatures. Here, the escape rate has a \textit{quantum origin} with $\frac{\sigma}{I_{\rm SW}} \approx(\frac{E_{\rm TH}}{ E_{\rm J}})^{(4/5)}$ ($E_{\rm J}$ Josephson energy) \cite{squids} up to temperatures of the order of $E_{\rm TH}$. 2) At higher temperatures, the process is driven by \textit{thermal activation} and scales with $\frac{\sigma}{I_{\rm SW}} \propto (\frac{\rm T}{\rm E_J})^{(2/3)}$ (see green dashed lines) \cite{squids}. 

\begin{figure}[!ht]
\includegraphics[width=1.0\linewidth]{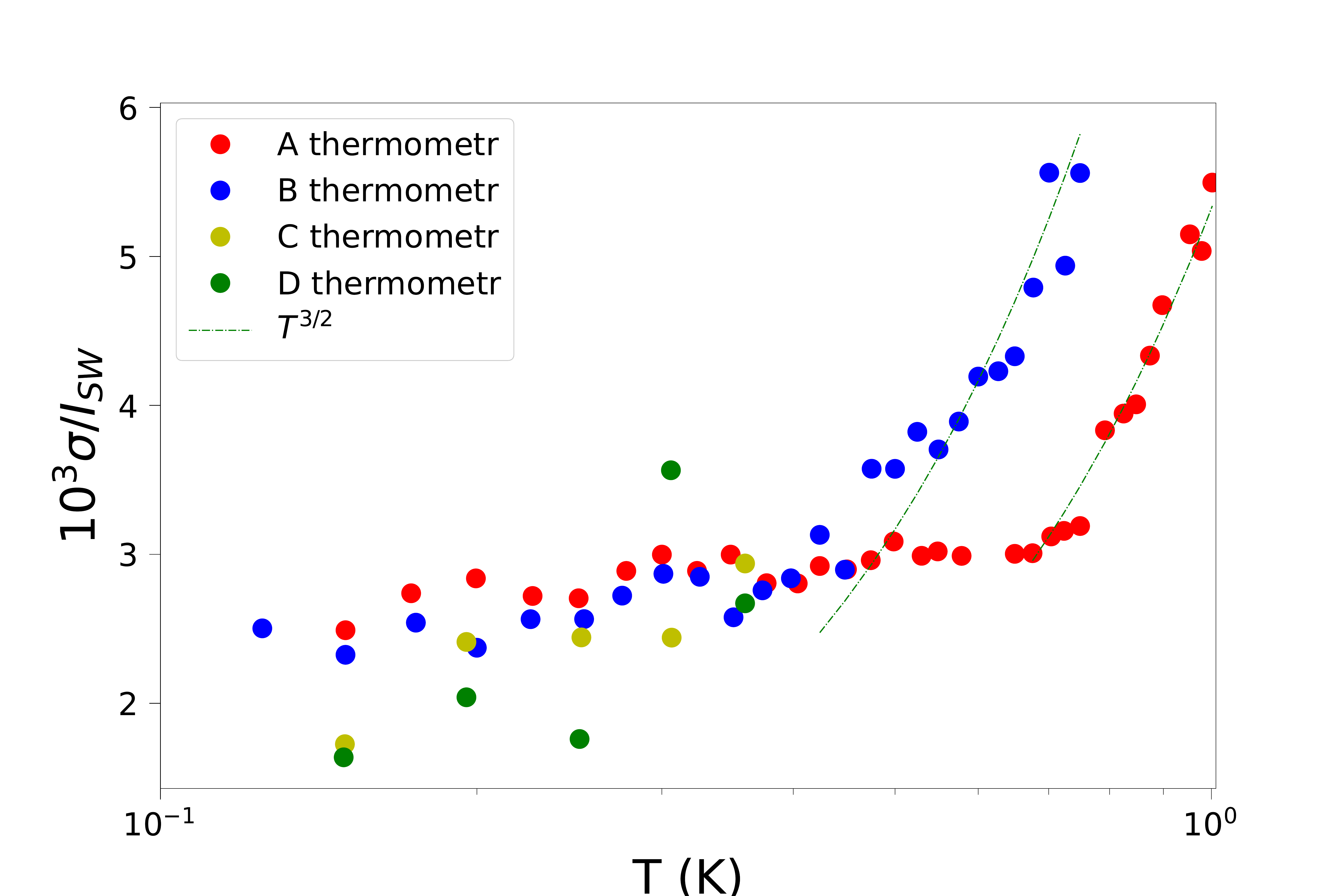}
\caption{The standard deviation $\sigma$ normalized by $I_{\rm SW}$ vs temperature for A (red), B (blue), C (yellow), and D (green) thermometers. The dashed green line is a theoretical fit assuming thermal activation (see SI III).}
\label{fig:fig3} 
\end{figure}

\section{Temperature resolution of thermometers}

The temperature resolution, $\epsilon_{\rm TR}$, of a thermometer is the minimum temperature that the thermometer can distinguish at a certain bath temperature, $T_{\rm bath}$, of the cryostat. We define this value as $\epsilon_{\rm TR}=-\frac{\Delta I}{dI_{\rm SW}/dT_{\rm bath}}$, where $\Delta I$ is full width at half maximum (FWHM) of measured switching current distribution at $T_{\rm bath}$. In Fig. S5  $\epsilon_{\rm TR}$ is plotted as a function of temperature for thermometers A and B (described in main text), and C,D (see SI.II).

\begin{figure}[!ht]
\includegraphics[width=1.0\linewidth]{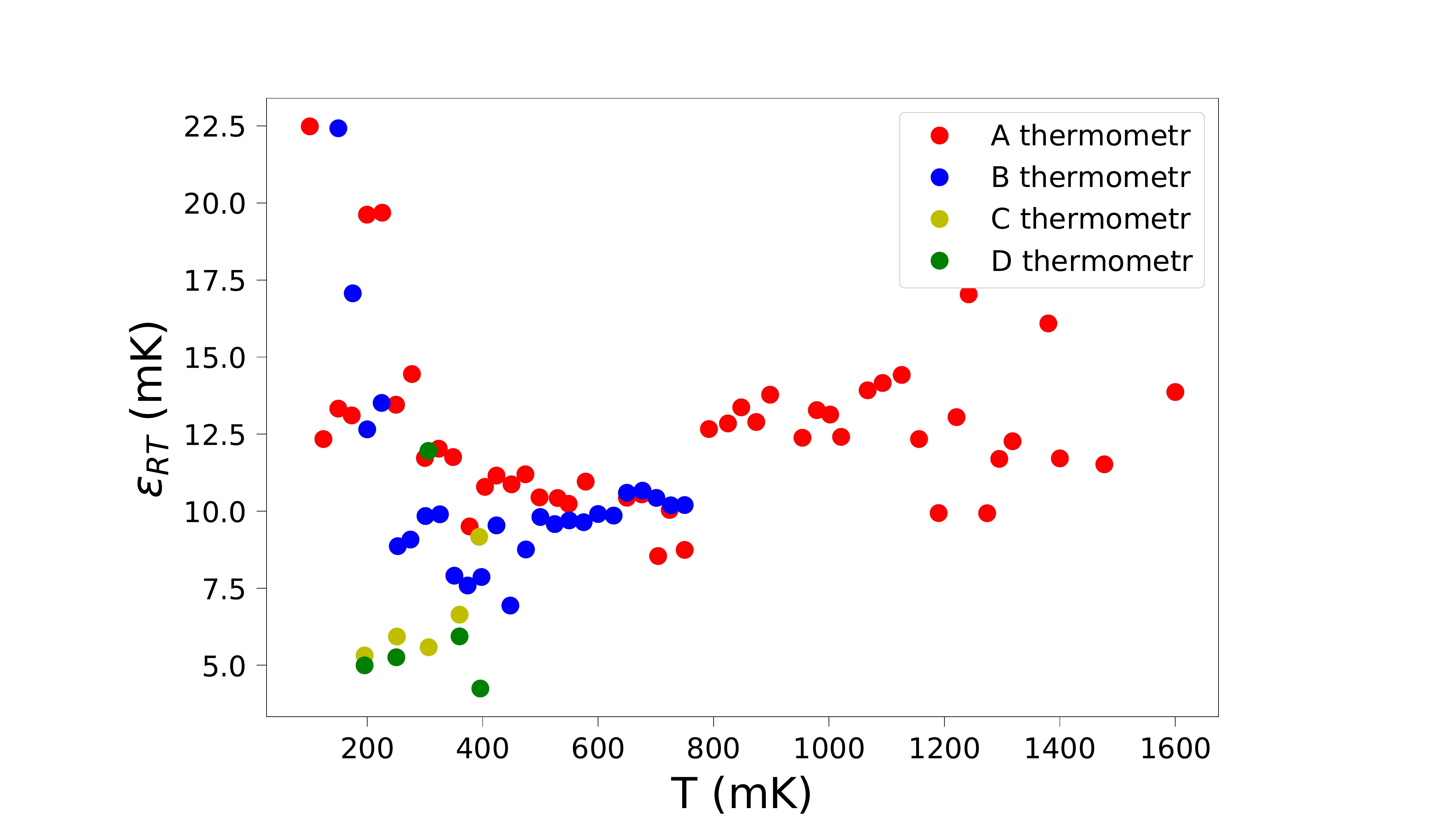}
\caption{The temperature resolution,$ \epsilon_{\rm TR}$ of thermometers A (red), B(blue), C (yellow), and D (green) vs. temperature.}
\label{fig:fig5} 
\end{figure}

Thermometers A and B, mentioned in main text, have a resolution between 10 and 15~mK which stays constant in the temperature range from 200~mK to 700~mK. We observe a slight decrease in resolution at the lower temperatures, which we attribute to the fact that the switching current is less dependent on temperature as $T$ approaches zero. The resolution of thermometers C and D is better than for A,B and reaches a value of 5~mK. Despite the fact that thermometers C and D perform better in the investigated temperature range (100 - 400~mK), for molecular devices we prefer to have thermometers with a higher $I_{\rm SW}$ and thus allowing to use them in a large temperature range.

\section{Other possible thermometers}

SNS thermometry can be readily implemented in  EMBJ devices as it does not require many additional lithographic steps and provides negligible access resistance, which is beneficial for conductance measurements. In this work, we used MoRe as a superconductor because of its high critical temperature. Below we present a list of parameters for previously reported SNS junctions that could be implemented for thermometry purposes. We limit this table to junctions that have Au as a normal metal because gold is the most used material to contact the molecule in single-molecule studies~\cite{ASU_chemprinciples}.

\begin{table}[h]
\centering
\begin{tabular}{ |c|c|c|c|c|c|c| } 
 \hline
SNS & $I_{\rm SW}$ ($\rm \mu A$) & $\rm R_N$ ($\Omega$) &$E_{\rm TH}$ ($\rm \mu eV$)& $\Delta$ ($\rm meV$)& $\frac{dI_{\rm SW}}{dT}^\ast$ ($\rm \mu A /K$)& Source  \\ 
 \hline
   Al-Au-Al& 18 & 4 &19& 0.17 &25   &\cite{Jung2011}\\
   Nb-Au-Nb& 178 & 1.7 &52.0& 1.23 &60  &\cite{Cecco}\\
   
 \hline
\end{tabular}
\caption{Parameters reported in literature for SNS junctions that can be implemented in molecular junctions for thermometry. $ I_{\rm SW}$ -switching current, $\rm R_N$ -normal resistance, $E_{\rm TH}$ - Thouless energy, $\Delta$ -superconducting gap, $\frac{dI_{\rm SW}}{dT}^\ast$ current resolution in a temperature range from the lowest measured temperature to 1 K.}
\label{table:4}
 
\end{table}

\bibliography{mycitations}